\documentclass[12pt]{article}

\def\sumint{\hbox{\small$\Sigma$}\kern-0.73em\int\kern.1em}

\usepackage{authblk}
\usepackage{scicite}
\usepackage{times}
\usepackage{graphicx}
\usepackage{xcolor}
\usepackage{array,multirow,graphicx}
\usepackage{multirow,hhline,graphicx,array}
\usepackage{rotating}
\usepackage{caption}
\usepackage{hyperref}
\usepackage{mathtools}
\usepackage[superscript,biblabel]{cite}
\newcolumntype{M}[1]{>{\centering\arraybackslash}m{#1}}

\newenvironment{sciabstract}{%
\begin{quote} \bf}
{\end{quote}}

\title{Isotopic effects in molecular attosecond photoelectron interferometry.}%Single-shot sideband correlation analysis for attosecond waveform synthesis}

\author[1,*]{Dominik Ertel}\author[1,2,*]{David Busto}\author[1]{Ioannis Makos}\author[1]{Marvin Schmoll}\author[3]{Jakub Benda}\author[1]{Hamed Ahmadi} \author[1]{Matteo Moioli}\author[4]{Fabio Frassetto}\author[4]{Luca Poletto} \author[5]{Claus Dieter Schr\"{o}ter}\author[5]{Thomas Pfeifer}\author[5]{Robert Moshammer}\author[3]{Zden\v{e}k Ma\v{s}\'{i}n}\author[6]{Serguei Patchkovskii}\author[1]{Giuseppe Sansone}
\affil[1]{Physikalisches Institut, Albert-Ludwigs-Universit\"{a}t Freiburg Hermann-Herder-Stra{\ss}e 3, 79104 Freiburg, Germany.}
\affil[2]{Department of Physics, Lund University, PO Box 118, SE-221 00 Lund, Sweden.}
\affil[3]{Institute of Theoretical Physics, Faculty of Mathematics and Physics, Charles University, V Holešovičkách 2, 180 00, Prague 8, Czech Republic.}
\affil[4]{Istituto di Fotonica e Nanotecnologie, CNR, 35131 Padova, Italy.}
\affil[5]{Max-Planck-Institut f\"{u}r Kernphysik, 69117 Heidelberg, Germany.}
\affil[6]{Max Born Institute, Max-Born-Str. 2A, D-12489 Berlin, Germany}
\affil[*] {These authors contributed equally to this work}

\date{}

\begin{document}


\baselineskip24pt


\maketitle


\begin{sciabstract}

Isotopic substitution in molecular systems can affect fundamental molecular properties including the energy position and spacing of electronic, vibrational and rotational levels, thus modifying the dynamics associated to their coherent superposition. In extreme ultraviolet spectroscopy, the photoelectron leaving the molecule after the absorption of a single photon can trigger an ultrafast nuclear motion in the cation, which can lead, eventually, to molecular fragmentation. This dynamics depends on the mass of the constituents of the cation, thus showing, in general, a significant isotopic dependence.
In time-resolved attosecond photoelectron interferometry, the absorption of the extreme ultraviolet photon is accompanied by the exchange of an additional quantum of energy (typically in the infrared spectral range) with the photoelectron-photoion system, offering the opportunity to investigate in time the influence of isotopic substitution on the characteristics of the photoionisation dynamics.
Here we show that attosecond photoelectron interferometry is sensitive to isotopic substitution by investigating the two-color photoionisation spectra measured in a mixture of methane (CH$_4$) and deuteromethane (CD$_4$). The isotopic dependence manifests itself in the modification of the amplitude and contrast of the oscillations of the photoelectron peaks generated in the two-color field with the two isotopologues. The observed effects are interpreted considering the differences in the time evolution of the nuclear autocorrelation functions of the two molecules.
\end{sciabstract}

Attosecond and strong-field spectroscopy have been widely used for the investigation of correlated electronic dynamics in atoms~\cite{CPC-Sansone-2012} and correlated electronic-nuclear dynamics in molecules~\cite{Nisoli2017}.
In atoms, strong field photoionisation can create a coherent superposition of electronic states of the cation, leading to electronic charge oscillations that have been resolved in time by attosecond time-resolved transient absorption~\cite{NATURE-Goulielmakis-2010}. In molecules, it was shown that the process of tunnelling ionisation launches a correlated electronic-nuclear dynamics~\cite{NATURE-Niikura-2003}, offering the possibility to use the electronic wave packet recolliding with the parent cation for imaging molecular structures~\cite{Science-Meckel-2008} and ultrafast nuclear dynamics~\cite{SCIENCE-Wouter-2016}. The latter is expected to play a crucial role for the observation of electronic processes in complex molecules~\cite{CPL-Astiaso-2017, PRL-Vacher-2017}.

An approach for studying the effect of nuclear motion on the correlated electronic-nuclear dynamics is the investigation of the response of molecules presenting the same structure and chemical composition, but with an isotopic substitution of one or more constituents. Under these conditions, the chemical properties of the systems, i.e. their electronic properties, are typically not significantly affected, giving the possibility to isolate the effect of the different timescale of the nuclear response in the isotopologues.
Isotopic effects in high-order harmonic generation (HHG) have been first predicted~\cite{PRL-Lein-2005} and then experimentally observed in hydrogen, deuterium and methane~\cite{Science-Baker-2006,JPB-Haessler-2009}. Nuclear-motion effects in HHG appear to be universal~\cite{PRL-Patchkovskii-2009}, and have been demonstrated in molecules as large as toluene~\cite{SCIREP-Austin-2021}. Recently, different decay times in the relaxation dynamics of the two isotopologues C$_2$H$_4$ and C$_2$D$_4$ have been observed in extreme ultraviolet (XUV)-infrared (IR) pump-probe experiments~\cite{JPCA-Vacher-2022}. Finally, the relevance of isotopic effects in methane (CH$_4$) and deuteromethane (CD$_4$) in the reorganization of the molecular structure of the cation after sudden ionisation~\cite{PCCP-Remacle-2021} was investigated theoretically~\cite{PRA-Serguei-2017}.

In attosecond photoelectron spectroscopy, the reconstruction of attosecond beating by interference of two-photon transitions (RABBIT) technique~\cite{Science-Paul-2001} has been widely used first for attosecond metrology applications and, later, for the investigation of attosecond electronic dynamics in all states of matter~\cite{RMP-Krausz-2009}. In particular, RABBIT measurements in molecules have evidenced the role played on photoionisation delays by shape-resonances~\cite{Huppert2015,Nandi2020} and by the emission direction of the outgoing photoelectron~\cite{Vos2018a,Ahmadi2022}. Moreover, the effect of the vibrational degrees of freedom and nuclear dynamics on the photoionisation phases was observed in nitrogen~\cite{PRA-Haessler-2009} and hydrogen~\cite{Cattaneo2018, Cattaneo2022}.

Methane and deuteromethane are ideal candidates for the investigation of coupled electronic-nuclear dynamics on an ultrashort timescale, due to the ultrafast non-adiabatic dynamics triggered by the photoionisation process. It manifests itself in the coupling between degenerate electronic and vibrational degrees of freedom, resulting in a reduced symmetry in the configuration of the molecular cation~\cite{Rabalais1971}. Exploiting the correlated dynamics of the electron-nuclear wave packet launched by tunnelling ionisation in an intense IR field, a different efficiency of the HHG process driven by the same intense IR field in these isotopologues was observed~\cite{Science-Baker-2006}.

Here, we characterise the effect of isotopic substitution in attosecond photoelectron interferometry by measuring, under the same experimental conditions, the photoelectron spectrograms generated in methane and its deuterated counterpart combining an attosecond pulse train and a synchronised IR field. The experimental data evidence an effect on the amplitude and on the contrast of oscillations of the photoelectron yield generated in the two-color photoionisation process. The experimental data can be interpreted considering the nuclear autocorrelation function of the two molecules, together with the different extension of the vibrational ground state.

\section*{\label{sec1} XUV-only photoelectron-photoion coincidence spectroscopy of methane and deuteromethane}
\begin{figure}
\centering \resizebox{1.0\hsize}{!}{\includegraphics{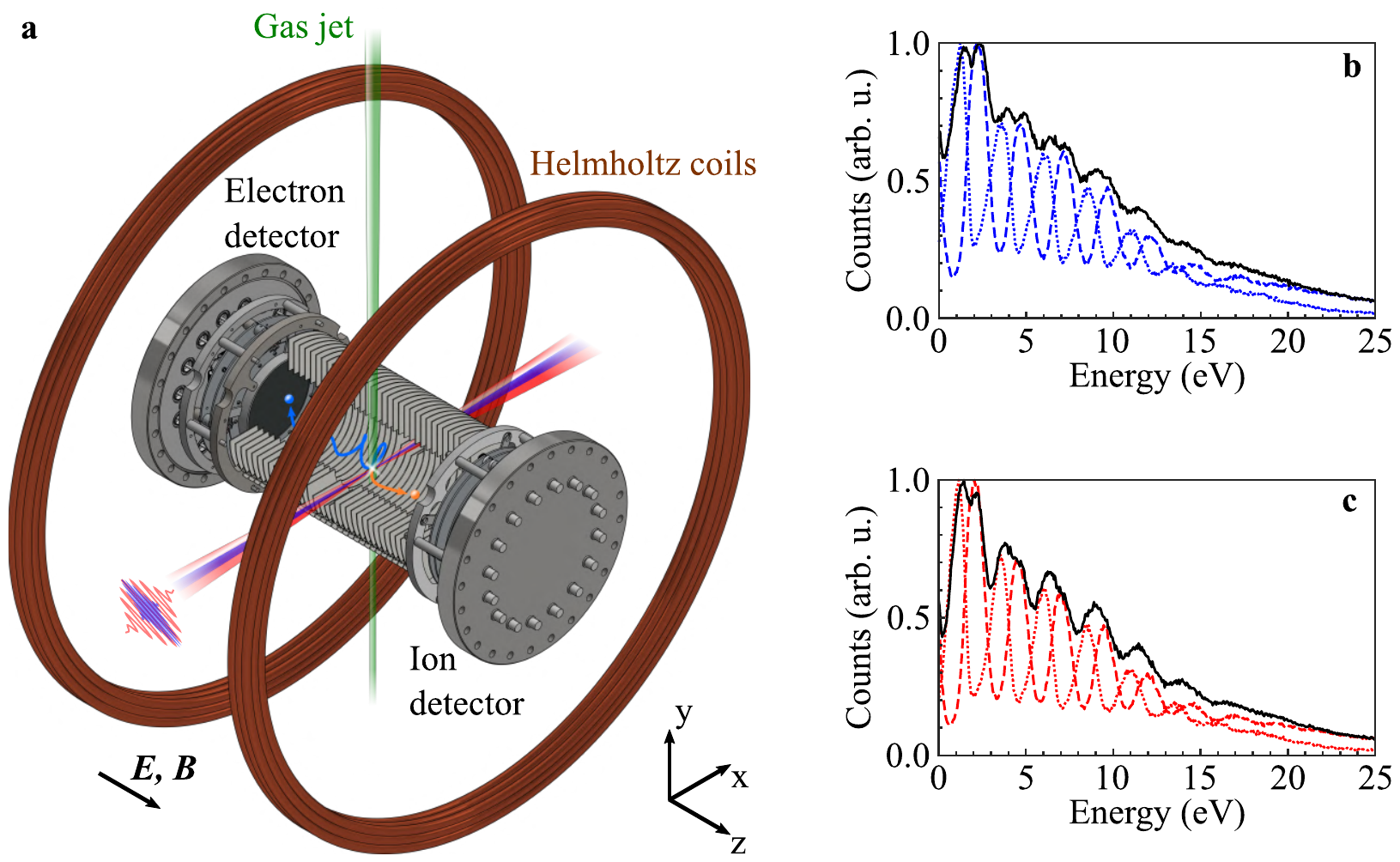}}
\caption{\textbf{Coincidence spectroscopy in a molecular isotopic mixture.} (a) Photoelectron-photoion coincidence spectrometer (reaction microscope) used in the experiment. (b) XUV-only photoelectron spectra measured in coincidence with the ionic fragments CH$_4^+$ (blue dashed line) and CH$_3^+$ (blue dotted line). (c) XUV-only photoelectron spectra measured in coincidence with the ionic fragments CD$_4+$ (red dashed line) and CD$_3^+$ (red dotted line). The black lines are the normalized spectra obtained as the sum of those measured in coincidence with the ions CH$_4^+$ and CH$_3^+$ (b), and CD$_4^+$ and CD$_3^+$ (c).}
\label{Fig1}
\end{figure}

We used a photoelectron-photoion coincidence spectrometer (reaction microscope, ReMi)~\cite{Moshammer1996,Ullrich1997,Dorner2000,Ullrich2003} to disentangle the photoelectron spectra resulting from the photoionisation of an equal mixture of CH$_4$ and CD$_4$ molecules (see Fig.~\ref{Fig1}a).
The ion spectra measured using the XUV attosecond pulse train alone are dominated by the ionic fragments CH$_4^+$-CH$_3^+$ and CD$_4^+$-CD$_3^+$~\cite{Chupka1968,Latimer1999}. Figure~\ref{Fig1}b and c show the total photoelectron spectra (black lines), and those measured in coincidence with the ions  CH$_4^+$ and CH$_3^+$ (blue dashed and dotted line, respectively) (Fig.~\ref{Fig1}b), and CD$_4^+$ and CD$_3^+$ (red dashed and dotted line, respectively) (Fig.~\ref{Fig1}c).
The channel-resolved photoelectron spectra present a clear harmonic structure. The results are consistent with the measurements obtained by photoelectron-photoion coincidence spectroscopy in combination with monochromatic XUV radiation~\cite{Field1995}. The capability to measure in coincidence the photoelectrons and the corresponding ions is fundamental for characterising the response of the two molecules under the same conditions, thus ruling out the effect of experimental instabilities, and to disentangle the contribution of the dissociating and non dissociating ionization channels.

The width of a single photoelectron peak is broader than the bandwidth of the corresponding XUV harmonic, estimated in about 200~meV (full-width at half maximum; FWHM). The broadening is due to the combination of the finite resolution of the photoelectron spectrometer (see Methods section) and the spectral width of the vibronic absorption band of methane and deuteromethane~\cite{Field1995}. We observe that the width of the single photoelectron peak for the XUV-only photoionisation process presents an isotopic dependence as shown in Extended Data Fig.~\ref{Fig1ED}a,b.
In particular, the FWHM of the single photoelectron peak measured in coincidence with the ionic channels CH$_4^+$ and CH$_3^+$ (blue solid lines in Extended Data Fig.~\ref{Fig1ED}a and b, respectively) are larger than the corresponding quantities measured in coincidence with the ions CD$_4^+$ and CD$_3^+$ (red dotted lines), respectively. This is demonstrated in Extended Data Fig.~\ref{Fig1ED}c,d, which report the difference of the FWHM ($\Delta$FWHM=FWHM$_{\mathrm{CH}^+_{3,4}}$-FWHM$_{\mathrm{CD}^+_{3,4}}$) for the photoelectron peaks measured in coincidence with ions associated to the CH$_4$ and CD$_4$ molecules.
As shown later, the broadening of the photoelectron peaks and the different widths can be explained considering the autocorrelation function of the two isotopologues.

\section*{\label{sec2} Attosecond photoelectron interferometry in methane and \\deuteromethane}
\begin{figure}
\centering \resizebox{1.0\hsize}{!}{\includegraphics{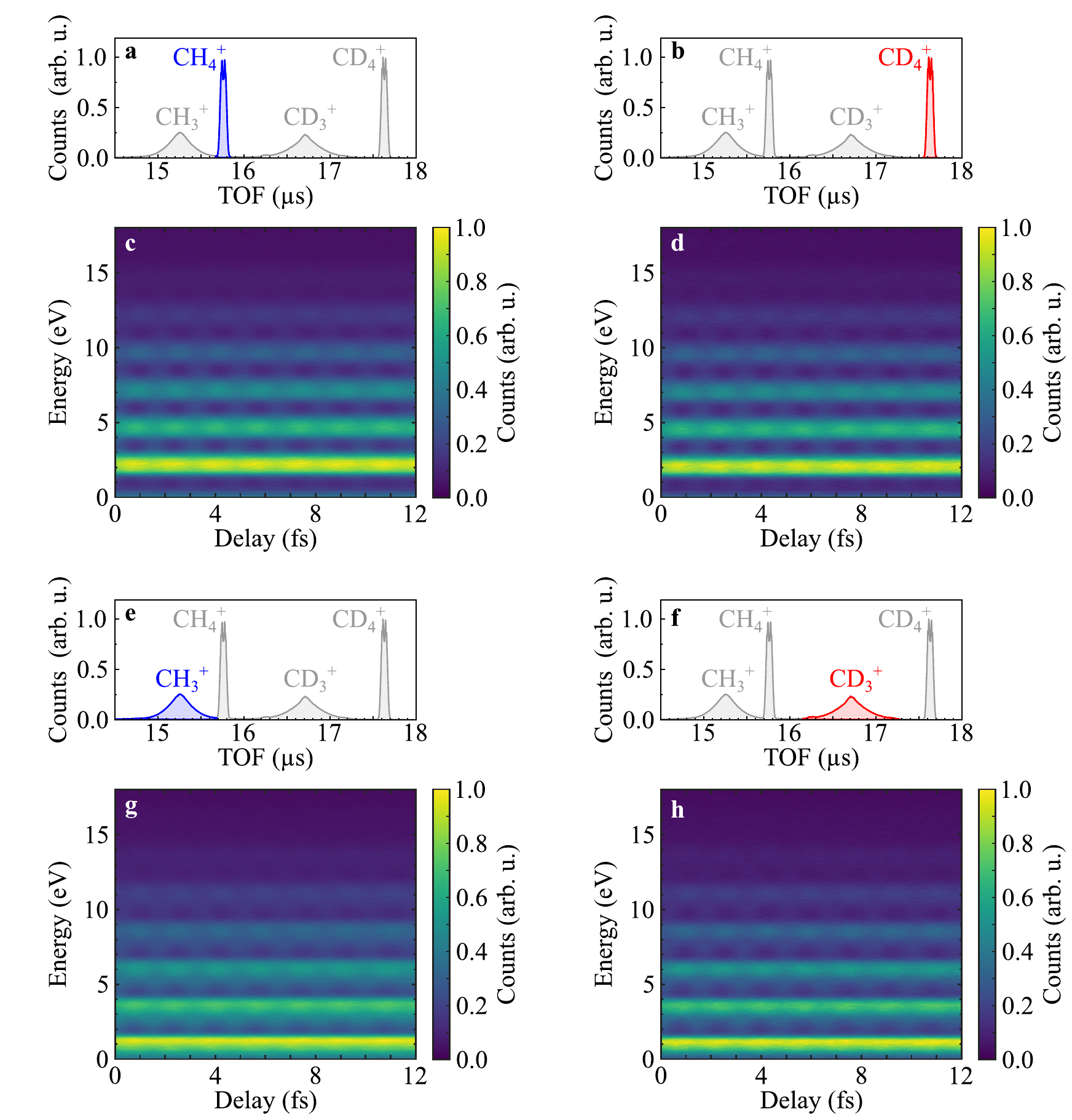}}
\caption{\textbf{Channel-resolved RABBIT spectrograms measured in methane and deuteromethane.} Ion time-of-flight spectra highlighting the ions CH$_4^+$ (a), CD$_4^+$ (b), CH$_3^+$ (e), and CD$_3^+$ (f). RABBIT traces obtained considering only photoelectron measured in coincidence with the ion CH$_4^+$ (c), CD$_4^+$ (d), CH$_3^+$ (g), and CD$_3^+$ (h).}
\label{Fig2}
\end{figure}

\begin{figure}
\centering \resizebox{0.6\hsize}{!}{\includegraphics{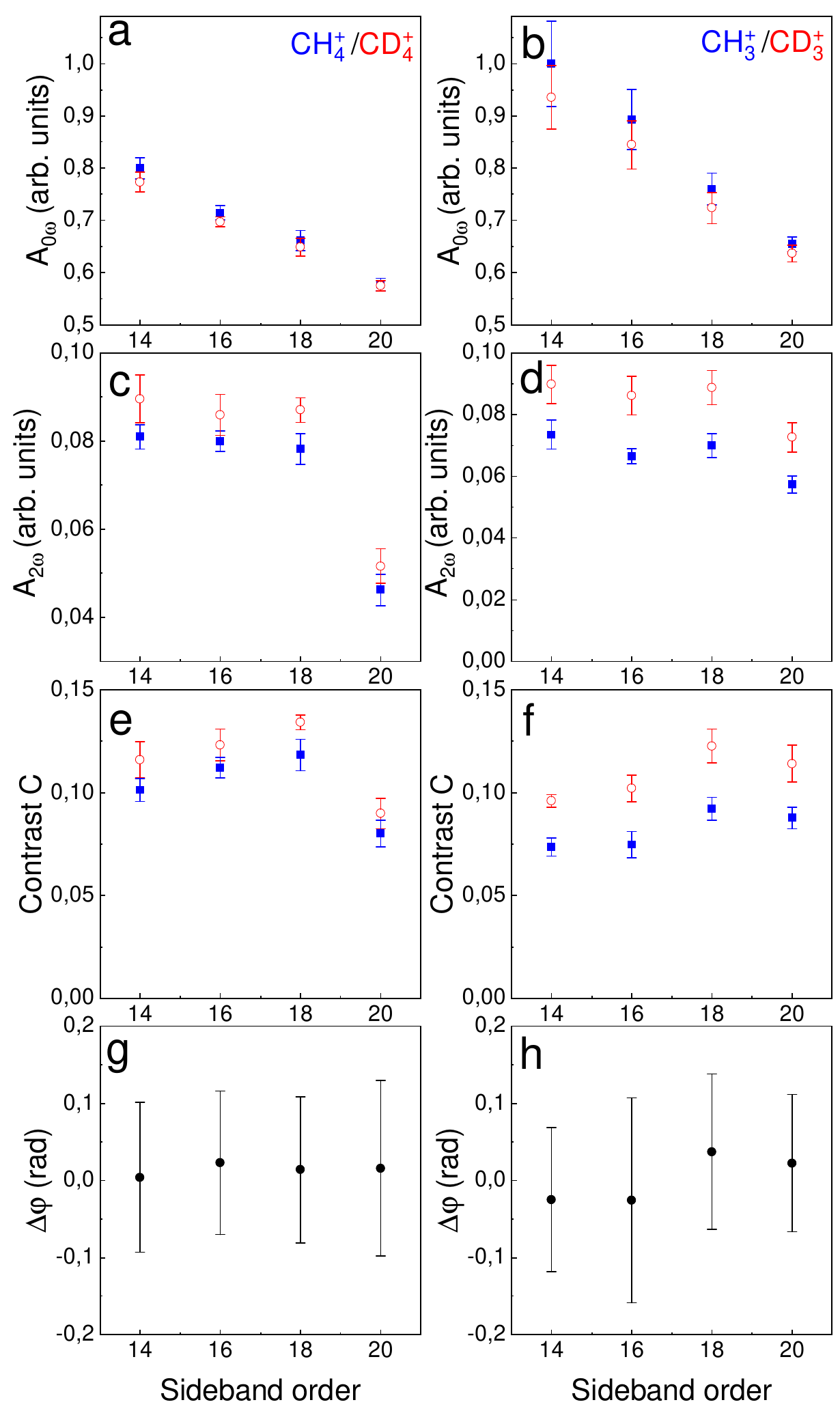}}
\caption{\textbf{Experimental isotopic dependence of sideband oscillations in methane and deuteromethane.}
Comparison between the amplitude offset $A_{0\omega}$ (a,b), amplitude $A_{2\omega}$ (c,d) and contrast $C=A_{2\omega}/A_{0\omega}$ (e,f) between the sidebands oscillations measured in coincidence with CH$_4^+$ (blue full square) and CD$_4^+$ (red open circle) (a,c,e), and with CH$_3^+$ (blue full square) and CD$_3^+$ (red open circle) (b,d,f) for different sideband orders. Phase difference $\Delta\varphi$ (g,h) (black full circle) of the phase offset $\varphi$ measured in coincidence with CH$_4^+$ and CD$_4^+$ (g), and with CH$_3^+$ and CD$_3^+$ (h) for different sideband orders. The error bars in panels a-f were obtained as standard deviation of the average values around the sideband maxima. For panels g and h, the error bars were evaluated as the error of the weighted mean calculated around the sideband maxima.}
\label{Fig3}
\end{figure}

When photoionisation takes place in the presence of a synchronised IR field, additional photoelectron peaks (sidebands) appear between the main photoelectron lines, due to the interference between two photoionisation pathways leading to the same final state. These involve the absorption of an XUV photon from consecutive harmonics and the absorption or emission of an additional IR photon by the outgoing photoelectron~\cite{Science-Paul-2001}. The yield of the sidebands measured in coincidence with the ions CH$_4^+$, CD$_4^+$, CH$_3^+$, and CD$_3^+$ (see ion TOF presented in Fig.~\ref{Fig2}a,b,e and f) clearly oscillate as a function of the relative delay $\Delta t$ between the attosecond pulse train and the IR field, as shown in Fig.~\ref{Fig2}c,d,g and h, respectively . The amplitude offset ($A_{0\omega}$), oscillation amplitude ($A_{2\omega}$) and phase offset ($\varphi$) of these oscillations were obtained by integrating the signal over an energy window of $\pm 300$~meV around the sideband maxima and by fitting the oscillations with the function:
\begin{equation}
    I^{SB}_{n}(\Delta t;E)=A_{0\omega}+A_{2\omega}\cos(2\omega\Delta t-\varphi),
    \label{Eq1}
\end{equation}
where $\omega$ indicates the frequency of the IR field driving the HHG process, $E$ is the kinetic energy of the photoelectron and $n$ is the sideband order. In Fig.~\ref{Fig3}, we present the comparison between the coefficients $A_{0\omega}$ (Fig.~\ref{Fig3}a,b), $A_{2\omega}$ (Fig.~\ref{Fig3}c,d) and the ratio $C=A_{2\omega}/A_{0\omega}$ (Fig.~\ref{Fig3}e,f), which corresponds to the contrast of the sideband oscillations, for the ionic channels resulting from photoionisation of CH$_4$ (blue full squares) and CD$_4$ (red open circles). Moreover, we also show the channel-resolved difference of the phase of the sideband oscillations $\Delta\varphi$ for the channels CH$_4^+$-CD$_4^+$ (Fig.~\ref{Fig3}g) and CH$_3^+$-CD$_3^+$ (Fig.~\ref{Fig3}h). A detailed description of the analysis followed to isolate the different terms is presented in the supplementary information (SI).

The amplitude offsets $A_{0\omega}$ are slighlty larger for the CH$_4^+$ channel than for CD$_4^+$ (Fig.~\ref{Fig3}a), as well as for CH$_3^+$ with respect to CD$_3^+$ (Fig.~\ref{Fig3}b). The amplitude of the oscillations $A_{2\omega}$ presents an opposite behaviour, as shown in Fig.~\ref{Fig3}c and d, with a clear isotopic difference between CH$_3^+$ and CD$_3^+$ (Fig.~\ref{Fig3}d). As a result, the contrast $C$ of the sideband oscillations turns out to be larger in the fragments originating from the CD$_4$ molecule (Fig.~\ref{Fig3}e,f). Interestingly, the differences of the phases $\varphi$ for the pairs CH$_4^+$-CD$_4^+$, and CH$_3^+$-CD$_3^+$ do not present a significant variation (see Fig.~\ref{Fig3}g and h). The experimental data indicate the presence of isotopic effects in the two-color photoionisation process that manifest themselves in the amplitude and contrast of the sideband oscillations, and not in the phase of the oscillations. The effect on the contrast suggests that the coherence properties of the two-color photoionisation~\cite{PRX-Bouchet-2020} process are affected by the isotopic substitution.

\section*{\label{sec3} Theoretical modelling and comparison with the experiment}
The RABBIT traces were simulated using perturbation theory, including up to second-order effect to take into account the exchange of two photons with the XUV-IR field (see Theoretical Model in Supplementary Information)~\cite{PRA-Patchkovskii-2022}. One- and two-photon electronic matrix elements were calculated using the stationary multiphoton molecular R-Matrix approach~\cite{SCIREP-Benda-2021,PRA-Benda-2022}. The effect of nuclear dynamics in the XUV-only and in the XUV-IR photoionisation processes was modelled using the vibronic auto-correlation function $A(\tau)$~\cite{ACR-Heller-1981}, where $\tau$ indicates the time elapsed since the absorption of the XUV photon. The autocorrelation function expresses the overlap between the time dependent nuclear wave packet $\chi(q,\tau)$ at the instant $\tau$ and the initial wavepacket $\chi(q,0)$, created in the cation at the time zero by the absorption of the energetic XUV photon ($q$ indicate the nuclear coordinates). A closely related approach was already adopted to describe nuclear-motion effects in HHG~\cite{PRL-Patchkovskii-2009,PRA-Serguei-2017,JCP-Mondal-2015,JCTC-Mondal-2014} and attosecond electron-hole migration~\cite{PRL-Vacher-2017,PRA-Arnold-2017,PCCP-Ruberti-2022}.
The modulation depth of the sideband as a function of the final electron momentum $p$ ($B(p)$) is proportional to the product of the electronic matrix element $M(p)$ and the Fourier transform of the nuclear autocorrelation $N(\varepsilon_p)$:
\begin{equation}\label{Eq2}
  B(p)\propto M(p)N(\varepsilon_p),
\end{equation}
where $\varepsilon_p$ is the vibrational energy and $M(p)$ is averaged over the ground state (zero-point-energy) vibrational function (see Supplementary Equations (7,8)).
%The matrix element $M(\mathbf{p})$ for a transition from the ground state to a continuum state with momentum $\mathbf{p}$ as
%\begin{equation}
%  M(\mathbf{p})=\langle \chi_0(q)|\hat{D}^{\dagger}_\mathbf{p}(q) \hat{D}_\mathbf{p}(q)|\chi_0(q)\rangle N_c(\varepsilon_p)=M^{\mathrm{el}}N_c(\varepsilon_p),
%  \label{Eq2}
%\end{equation}
%where $\hat{D}_\mathbf{p}$ indicates the standard electronic matrix operator for two-photon absorption from the ground state to a continuum state with final momentum $\mathbf{p}$, $\chi_0(q)$ is the ground-state vibrational wave function of the neutral molecule dependent on the nuclear coordinates $q$, and $N_c(\varepsilon_p)$ is the Fourier transform of the autocorrelation function $A_c(t_d)$ and depends on the vibrational energy $\varepsilon_p$ (see Theoretical Model in Supplementary Information). The matrix element $M(\mathbf{p})$ takes into account also averaging effects over the ground-state vibrational wave functions of the two isotopes.}

\begin{figure}
\centering \resizebox{0.9\hsize}{!}{\includegraphics{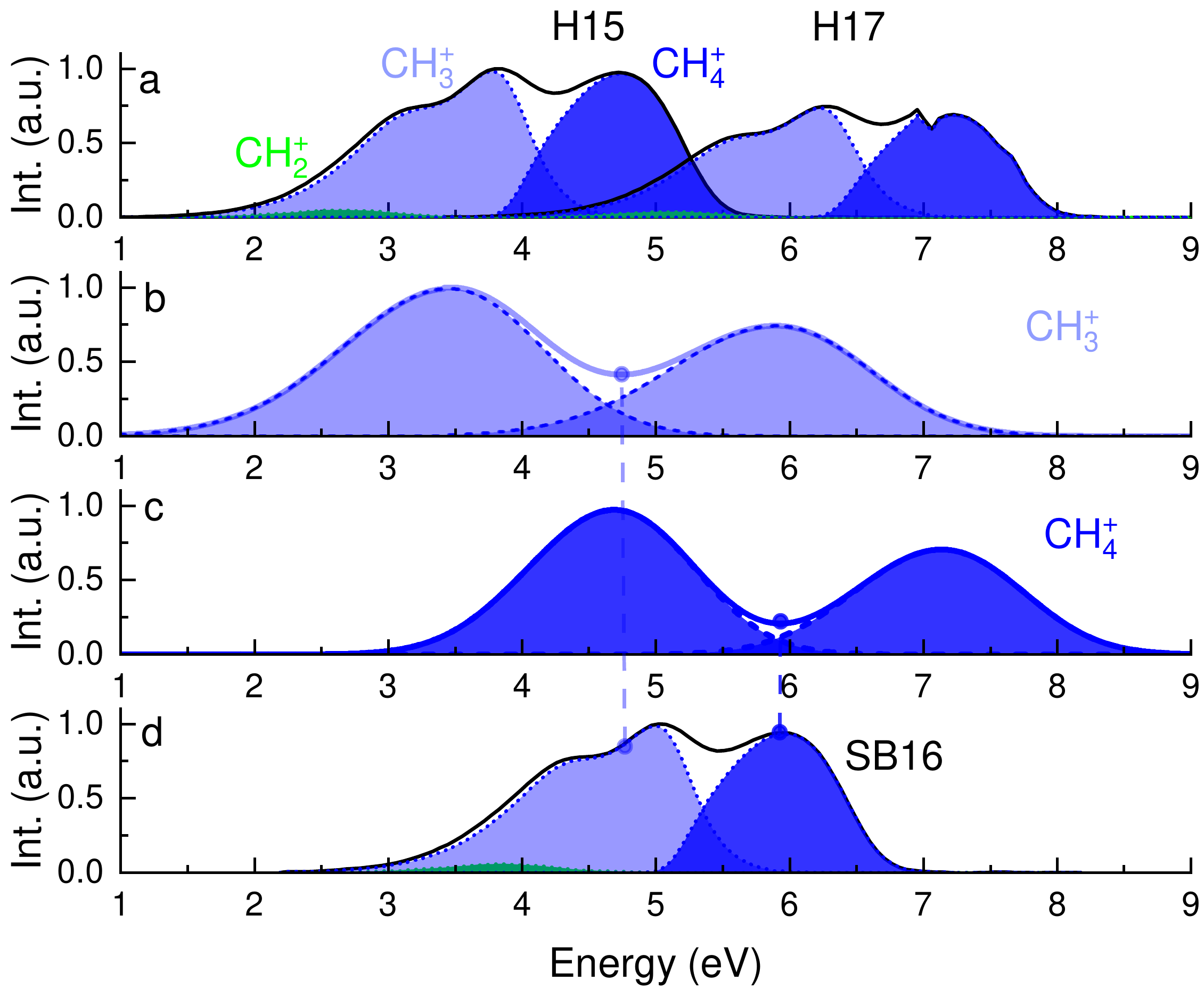}}
\caption{\textbf{Simulated XUV-only and two-color photoelectron spectra.} a) Main photoelectron lines (black solid line) determined by the absorption of a single XUV-photon of the harmonics H15 and H17 in CH$_4$. The shaded areas indicate the photoelectron spectra associated to the ionic channels CH$_2^+$ (green), CH$_3^+$ (light-blue) and CH$_4^+$ (blue). Simulated XUV-only photoelectron spectra (full solid line) considering only the contributions of the photoelectrons associated to CH$_3^+$ (b) and CH$_4^+$ (c) and the harmonics H15 and H17. The dashed lines and shaded areas indicate the simulated spectra obtained considering a single harmonic and the convolution with the response function of the ReMi photoelectron spectrometer. d) Profile of the photoelectron sideband SB16 (black line) and its decomposition in the contributions  associated to the ionic channels CH$_2^+$ (green), CH$_3^+$ (light-blue) and CH$_4^+$ (blue). The vertical dashed lines indicate the positions of the minima of the XUV-only photoelectron spectra (light-blue for CH$_3^+$ and blue for CH$_4^+$).}
\label{Fig4}
\end{figure}

The simulated photoelectron spectra due to the absorption of a single XUV photon from the harmonic-order $n=15$ and $n=17$ are presented in Fig.~\ref{Fig4}a (black lines). The overall spectral shape is dominated by the shape of the function $N(E)$ (with the photoelectron kinetic energy $E=-\varepsilon_p+n\hbar\omega$) and matches the photoelectron spectra measured with monochromatic XUV radiation well~\cite{Field1995} (see Extended Data Fig.~\ref{Fig2ED}a). The overall photoelectron spectra can be decomposed into three contributions corresponding to the dissociative (CH$_2^+$ and CH$_3^+$) and non-dissociative (CH$_4^+$) channels~\cite{Mahan1982}, respectively.
We used filters based on the measured branching ratios~\cite{Field1995} to isolate these three contributions (see Extended Data Fig.~\ref{Fig2ED}b), thus reproducing the ion-resolved photoelectron detection of the experiment. The result is represented by the shaded areas shown in Fig.~\ref{Fig4}a for the channel CH$_2^+$ (green area), CH$_3^+$ (light-blue area), and CH$_4^+$ (blue area). The XUV-only spectrum is constructed as the sum of the contributions of the different harmonics (H15 and H17 in Fig.~\ref{Fig4}), convoluted with the response function of the photoelectron spectrometer, as shown for the XUV-only spectrum associated with the ions CH$_3^+$ (Fig.~\ref{Fig4}b) and CH$_4^+$ (Fig.~\ref{Fig4}c), respectively. A similar procedure, considering the autocorrelation function for CD$_4$ was used for simulating the XUV spectra associated with the ions CD$^+_3$ and CD$^+_4$.

The simulated XUV-only spectra reproduce the experimental photoelectron spectra accurately (see Extended Data Fig.~\ref{Fig3ED}a,b). The difference of the spectral widths of the simulated peaks match qualitatively those of the experimental photoelectron lines measured in coincidence with the ions CH$_4^+$-CD$_4^+$ and CH$_3^+$-CD$_3^+$, as shown by the comparison between Extended Data Fig.~\ref{Fig1ED}c,d and Extended Data Fig.~\ref{Fig3ED}c,d, respectively. In particular, the larger difference in the dissociating channels (CH$_3^+$-CD$_3^+$; Extended Data Fig.~\ref{Fig3ED}d) with respect to the non-dissociating one (CH$_4^+$-CD$_4^+$; Extended Data Fig.~\ref{Fig3ED}c) is reproduced in the simulations.

%A similar agreement holds true also for the ions emitted starting from the isotope CD$_4$ (see Extended Data Fig.~\ref{Fig4ED}c,d).
%The validity of the decomposition is further supported by considering the differences in the widths of the simulated XUV-only harmonics and those estimated from the single photoelectron peaks measured in coincidence with the ions using the expression:
%\begin{equation}\label{Eq2}
%  \Delta w_{\mathrm{exp}}=(w_{\mathrm{CH}_4^+}+w_{\mathrm{CH}_3^+})-(w_{\mathrm{CD}_4^+}+w_{\mathrm{CD}_3^+})
%\end{equation}
%The average of the different sidebands $\Delta w_{\mathrm{exp}}=0.36\pm0.05$~eV is in excellent agreement with the value extracted from the simulations $\Delta w_{\mathrm{theo}}=0.30\pm0.04$~eV.

The good agreement between the channel-resolved experimental data measured in the XUV-only case and the simulated spectra allows one to extend the approach to the two-color photoionisation process for both isotopologues, including also the contribution of the sideband photoelectron peaks. Figure~\ref{Fig4}d presents the simulated spectrum for the sideband SB16, together with the filtered contributions corresponding to the ionic channels CH$_2^+$ (green area), CH$_3^+$ (light-blue area), and CH$_4^+$ (blue area). As expected, the maxima of the sidebands associated to the CH$_3^+$ (light-blue) and CH$_4^+$ (blue) channels correspond to the minima between consecutive harmonics of the corresponding XUV spectrum, as indicated by the vertical dashed lines in Fig.~\ref{Fig4}b,c,d. We constructed the complete RABBIT traces generated by the two-color process as the incoherent sum of the high (low) energy contributions of the main photoelectron lines and of the sideband peaks for the ionic channels CH$_4^+$ (CH$_3^+$). The incoherent sum is justified by the observation that the one- and two-photon pathways leading to the same final photoelectron energy, will access different vibrational states of the ion, therefore suppressing the interference between the two paths. The validity of this approach is supported by the good agreement between the delay ($\Delta t$) integrated experimental and simulated RABBIT traces (see Extended Data Fig.~\ref{Fig4ED}).
%the different number of photons involved in the photoionisation process as shown in ref.~\cite{PRR-Fuchs-2021}. Indeed, averaging over time delays between XUV and IR pulses -- corresponding to averaging over
% the random-varying carrier-envelope phase of the pulses used in the experiment leads to the washing out of the coherent interference term.
%The simulated channel-resolved photoelectron spectra averaged over the XUV-IR relative delay $\Delta t$ well match the corresponding experimental quantities (Extended Data Fig.~\ref{Fig4ED}).
%In the simulations, we do not account for the oscillations in the harmonic yields, since it would correspond to a third order process according to perturbation theory, while our approach is limited to effects up to the second order.

\begin{figure}
\centering \resizebox{0.6\hsize}{!}{\includegraphics{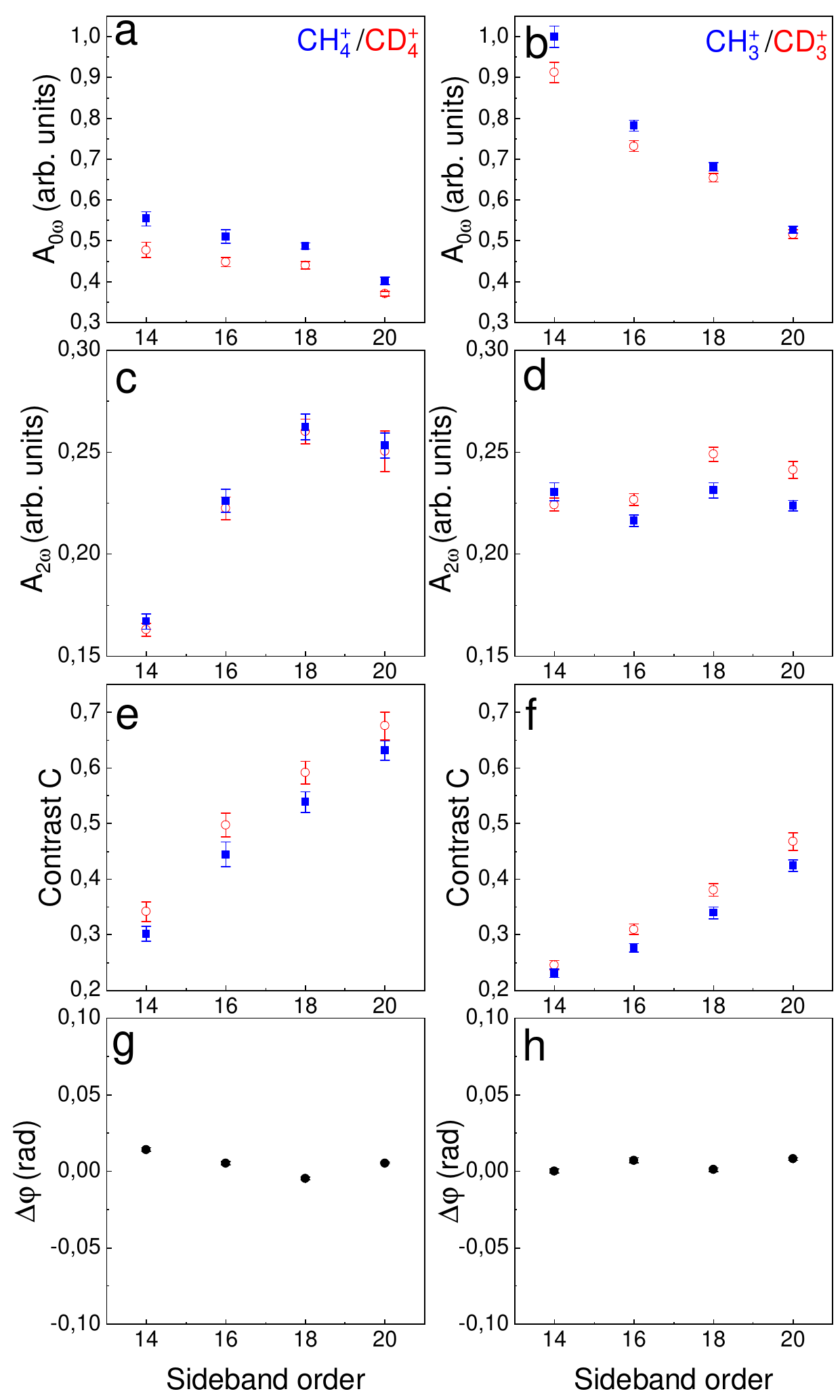}}
\caption{\textbf{Simulated isotopic dependence of sideband oscillations in methane and deuteromethane.} Comparison between the amplitude offset $A_{0\omega}$ (a,b), amplitude $A_{2\omega}$ (c,d) and contrast $C=A_{2\omega}/A_{0\omega}$ (e,f) between the sidebands oscillations associated with CH$_4^+$ (blue full square) and CD$_4^+$ (red open circle) (a,c,e), and with CH$_3^+$ (blue full square) and CD$_3^+$ (red open circle) (b,d,f) for different sideband orders. Phase difference $\Delta\varphi$ (g,h) (black full circle) of the phase offset $\varphi$ associated with CH$_4^+$ and CD$_4^+$ (g), and with CH$_3^+$ and CD$_3^+$ (h) for different sideband orders. The error bars in panels a-f were obtained as standard deviation of the average values around the sideband maxima. For panels g and h, the error bars were evaluated as the error of the weighted mean calculated around the sideband maxima.}
\label{Fig5}
\end{figure}

The simulated RABBIT spectrograms (reported in Extended Data Fig.~\ref{Fig5ED}) were analysed according to Eq.~\ref{Eq1} to extract the coefficients $A_{0\omega}$, $A_{2\omega}$, and $\varphi$. The results are presented in Fig.~\ref{Fig5} and are in close qualitative agreement with the experimental findings. The model reproduces the differences in $A_{0\omega}$ between the ionic channels observed in the experiment (see Fig.~\ref{Fig5}a,b). In particular, the amplitude offsets $A_{0\omega}$ of the sidebands associated with the ionic fragments CH$_4^+$ (Fig.~\ref{Fig5}a) and CH$_3^+$ (Fig.~\ref{Fig5}b) are larger than the corresponding quantities in CD$_4^+$ and CD$_3^+$. Furthermore, the model reproduces the trends in the amplitudes $A_{2\omega}$, with a negligible difference for the pair CH$_4^+$-CD$_4^+$ (Fig.~\ref{Fig5}c) and a significant one for the ionic fragments CH$_3^+$-CD$_3^+$ (Fig.~\ref{Fig5}d). These observations are in good agreement with the experimental results shown in Fig.~\ref{Fig4}c,d.
As in the experiment, the final result is an increased contrast $C$ of the oscillations for the sidebands originating from the CD$_4$ molecule (Fig.~\ref{Fig5}e,f), which is particularly pronounced in the comparison CD$_3^+$-CH$_3^+$ (Fig.~\ref{Fig5}f). In general, the amplitude of the oscillations at frequency $2\omega$ and the contrast estimated from the numerical model are higher than those measured in the experiment. We attribute these differences to the effect of averaging over the interaction volume and small distortions in the electric and magnetic fields used in the photoelectron spectrometer, which might lead to a partial smearing out of the sideband oscillations in the experimental data.

Finally, the model also reproduces the absence of significant variation in the phase difference between homologue isotopic species (CH$_4^+$-CD$_4^+$ and CH$_3^+$-CD$_3^+$) (Fig.~\ref{Fig5}g,h).

\section*{Interpretation}
The isotopic differences observed in the experiment can be interpreted considering the autocorrelation function $A(\tau)$ and its Fourier Transform $N(\varepsilon_p)$ in the spectral domain, together with the decomposition of the photoelectron spectra in the high- and low-energy components presented in Fig.~\ref{Fig4} (hereafter, we will use the subscripts $\mathrm{H}$ and $\mathrm{D}$ to indicate quantities related to the isotopologues CH$_4$ and CD$_4$, respectively).

\begin{figure}
\centering \resizebox{1.0\hsize}{!}{\includegraphics{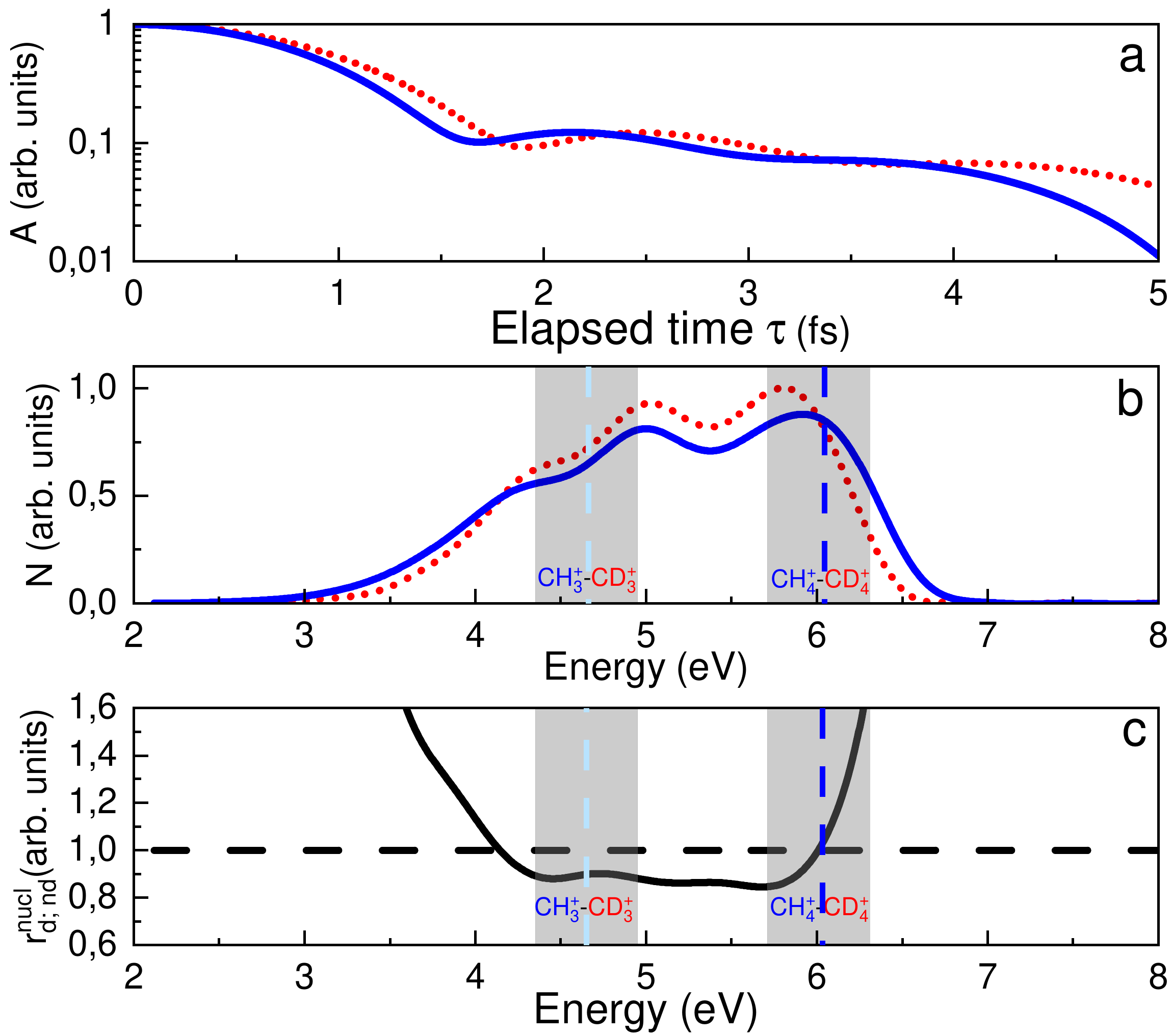}}
\caption{\textbf{Interpretation of isotopic dependence in terms of nuclear autocorrelation function.} a) Nuclear autocorrelation functions $A$ for CH$_4$ (blue full line) and CD$_4$ (red dotted line). %Train of attosecond pulses with a temporal width of 200~as (gray shaded area). b) Product of the train of attosecond pulses and of the autocorrelation functions for CH$_4$ (blue shaded areas) and CD$_4$ (red shaded areas).
b) Fourier transform of the autocorrelation function for CH$_4$ (blue full line) and CD$_4$ (red dotted line) shifted in the energy region of the SB16. c) Ratio of the Fourier transform of the autocorrelation functions of CH$_4$ and CD$_4$. The horizontal dashed line indicates the value $\mathrm{r^{nuc}_{d,nd}}=1$. The vertical dashed lines correspond to the energy position of the center of the sideband photoelectrons associated to the ionic channels CH$_3^+$-CD$_3^+$ (light-blue) and CH$_4^+$-CD$_4^+$ (blue). The grey shaded regions indicate the energy intervals corresponding to the integration intervals ($\pm 300$~meV) of the sideband photoelectrons associated to the ionic channels CH$_3^+$-CD$_3^+$ (left area) and CH$_4^+$-CD$_4^+$ (right area) (see also vertical lines in Fig.~\ref{Fig4}d).}
\label{Fig6}
\end{figure}

Due to its lighter mass, the dynamics triggered in the cation CH$_4^+$ by the absorption of an XUV photon is faster than in CD$_4^+$, leading to a faster decay in the very first few femtoseconds of the autocorrelation function $A_\mathrm{H}$ with respect to $A_\mathrm{D}$ (blue solid line and red dotted line, respectively in Fig.~\ref{Fig6}a).
%The autocorrelation functions (blue solid line for CH$_4^+$ and red dotted line for CD$_4^+$ in Fig.~\ref{Fig6}a) are characterized by a different decay time and also by local minima and maxima occurring at different times, depending on the specific isotope. %In the temporal domain, the dipole moment depends on the product of the attosecond pulse train (shown schematically in Fig.~\ref{Fig6}a as shaded grey areas with a FWHM duration of 200~as) and the autocorrelation function. We can therefore introduce the physical picture of an \emph{effective} attosecond pulse train experienced by the molecule, given by the product of the XUV attosecond pulse train and the autocorrelation function that present different temporal characteristics for CH$_4$ and CD$_4$, as indicated by the blue and red shaded areas in Fig.~\ref{Fig6}b, respectively. In general, the fast nuclear dynamics occurring after the initial absorption of an XUV photon determines a short time window (only few fs) over which the action of consecutive attosecond pulses in the train coherently builds up. This limits the effective duration of the waveform in the temporal domain to only a few attosecond pulses~\cite{PNAS-Cheng-2020}. The effect is more pronounced in CH$_4$ due to the faster decay of the autocorrelation function, which determined also a reduced temporal width of the effective attosecond pulse train (see differences in the effective attosecond pulse trains represented by the blue and red shaded areas in Fig.~\ref{Fig6}b).

In the energy domain, the difference between the decay times determines a different broadening of their Fourier transforms, resulting in the function $N_\mathrm{H}$ (blue solid line) being broader than $N_\mathrm{D}$ (red dotted line) both at low and at high energies, as shown in Fig.~\ref{Fig6}b.
%The shape of the function $N$ determines the evolution of photoelectron spectra simulated for the XUV-only (see Fig.~\ref{Fig4}a) and for the XUV-IR cases (see Fig.~\ref{Fig4}d).

As shown in Extended Data Fig.~\ref{Fig2ED}, the shape of the functions $N_\mathrm{H,D}$ reproduces well the photoelectron spectra generated by each harmonic of the XUV spectrum. Moreover, the single photoelectron peak can be decomposed in a high energy part associated to the ionic channels CH$_4^+$-CD$_4^+$ and low energy contribution associated (mainly) to CH$_3^+$-CD$_3^+$ (see Fig.~\ref{Fig4}).
The combination of these observations indicates that the larger broadening of the photoelectron peaks measured in coincidence with the ionic channel CH$_4^+$ with respect to CD$_4^+$ is due to the larger width of the function $N_\mathrm{H}$ at high energies (see Extended Data Fig.~\ref{Fig1ED}a,c). Similarly, the larger broadening of the photoelectron peaks measured in coincidence with the ionic channel CH$_3^+$ is due to the differences of the functions $N_\mathrm{H}$ and $N_\mathrm{D}$ for the two isotopologues at low energies (see Extended Data Fig.~\ref{Fig1ED}b,d).

In the two-color photoionisation measurements, the differences observed in the amplitudes $A_{0\omega}$ can be also explained by the different spectral width of the functions $N_{\mathrm{H}}$ and $N_{\mathrm{D}}$.
The larger $A_{0\omega}$ component in CH$_4^+$ and CH$_3^+$ with respect to their deuterated counterparts can be ascribed to the larger width of the XUV-only spectra, which determines a larger contribution of the photoelectron released by the absorption of a single XUV photon to the energy intervals between consecutive harmonics, where the two-photon (XUV-IR) sideband signal is located (see Fig.~\ref{Fig4}d).

For the interpretation of the evolution in the $A_{2\omega}$ components, we observe that the difference between the non-dissociating channels CH$_4^+$-CD$_4^+$ is significantly smaller than for the dissociating channels CH$_3^+$-CD$_3^+$, both in the experiment (Fig.~\ref{Fig3}c,d) and in the simulations (Fig.~\ref{Fig5}c,d). We can quantify this difference introducing the ratio between the amplitude of the sideband oscillations for the dissociating ($\mathrm{r_d}=A_{2\omega}(\mathrm{CH_3^+})/A_{2\omega}(\mathrm{CD_3^+})$) and non-dissociating ($\mathrm{r_{nd}}=A_{2\omega}(\mathrm{CH_4^+})/A_{2\omega}(\mathrm{CD_4^+})$) channels. The ratio $\mathrm{r_{nd}}$ is very close to one and larger than $\mathrm{r_d}$ both in the experiment and simulations, as shown in Extended Data Tab.~\ref{Table1}and Extended Data Fig.~\ref{Fig6ED}a,b.

%in the central energy region (-1~eV$<\varepsilon_p<0.5$~eV), the function $N$ for CH$_4$ is larger than that of CD$_4$. The
In our numerical model the amplitude of the oscillations of the sideband intensities $A_{2\omega}$ is proportional to the modulus of the Fourier transform of the modulation depth $B(p)$ at the frequency $2\omega$ (see Eq.~\ref{Eq2}). This can be expressed as the product of the modulus of an electronic term ($|M_{\mathrm{H,D}}|$) (shown in Extended Data Fig.~\ref{Fig7ED}a) and a nuclear contribution given by the function $N_{\mathrm{H,D}}$. (see Supplementary Information). As a result, the ratios $\mathrm{r_d}$ and $\mathrm{r_{nd}}$ can be expressed as the product of a ratio for the electronic contribution ($\mathrm{r^{el}_{d,nd}}$) and one for the nuclear part ($\mathrm{r^{nucl}_{d,nd}}$):
\begin{eqnarray}\label{Eq3}
  %r_{d,nd}&=&\frac{A_{2\omega}(\mathrm{CH_{3,4}^+})}{A_{2\omega}(\mathrm{CD_{3,4}^+})}=\underbrace{\frac{M^{\mathrm{el}}(\mathrm{CH_{3,4}^+})}{M^{\mathrm{el}}(\mathrm{CD_{3,4}^+})}}_\text{$r^{\mathrm{el}}_{d,nd}$}\underbrace{\frac{N_c(\mathrm{CH_{3,4}^+})}{N_c(\mathrm{CD_{3,4}^+})}}_\text{$r^{\mathrm{nucl}}_{d,nd}$}
  \mathrm{r_{d,nd}}&=&\frac{|{\cal F}\left[B_{\mathrm{H}}\right]_{2\omega}|}{|{\cal F}\left[B_{\mathrm{D}}\right]_{2\omega}|}=\underbrace{\frac{|M_{\mathrm{H}}|}{|M_{\mathrm{D}}|}}_\text{$\mathrm{r^{el}_{d,nd}}$}\underbrace{\frac{N_{\mathrm{H}}}{N_{\mathrm{D}}}}_\text{$\mathrm{r^{nucl}_{d,nd}}$}
  %\underbrace{_\text{$r^{\mathrm{el}}_{d,nd}$}\underbrace{\frac{N_c(\mathrm{CH_{3,4}^+})}{N_c(\mathrm{CD_{3,4}^+})}}_\text{$r^{\mathrm{nucl}}_{d,nd}$}
\end{eqnarray}
In the model, the ratio of the electronic terms depends only weakly on the dissociating or non-dissociating channel, through its energy dependence. For example, they assume the values $\mathrm{r^{el}_{d}}=1.052$ and  $\mathrm{r^{el}_{nd}}=1.042$ at the center of the sideband SB16 for the dissociating and non dissociating channels, respectively (see Extended Data Fig.~\ref{Fig7ED}b). The values for the other sidebands are reported in Table~\ref{Table1}, indicating that the electronic part of the amplitude of the sideband oscillation in CH$_4$ is larger than in CD$_4$. This ratio is modified by the nuclear contribution as shown in Fig.~\ref{Fig6}b; at the center of the region corresponding to the sidebands measured in coincidence with the ion CH$_3^+$-CD$_3^+$ (indicated by the light blue vertical dashed line) the function $N_\mathrm{H}$ (blue solid line) is smaller than $N_\mathrm{D}$ (red dotted line), leading to a ratio of the nuclear contribution $\mathrm{r^{nucl}_{d}}$ lower than one, which compensates the ratio of the electronic part $\mathrm{r^{el}_{d}}$ . For the non-dissociating channel CH$_4^+$-CD$_4^+$ (vertical blue dashed line), the ratio of the two curves is close to one.

We note that both in the experiment and in the simulations the sideband photoelectrons are integrated over an energy region (grey shaded areas in Fig.~\ref{Fig6}b,c) of about $\pm 300$ meV around the sideband maxima. In the case of the dissociation channel (CH$_3^+$ and CD$_3^+$) (grey shaded area on the left handside), the function $N_\mathrm{H}$ remains smaller than $N_\mathrm{D}$ over the entire interval, while for the non dissociating channel (CH$_4^+$ and CD$_4^+$;  grey shaded area on the right-hand side), the two curves cross.  The nuclear contribution $\mathrm{r^{nucl}_{d,nd}}$ can be approximated as the ratio of the integrals of the curves $N_{\mathrm{H,D}}$ over the energy intervals corresponding to the sidebands, returning the values $\mathrm{r_{d}^{nucl}}=0.894$ and $\mathrm{r_{nd}^{nucl}}=1.017$. The results are summarized in Extended Data Table~\ref{Table1}. The factorisation of the ratio $\mathrm{r_{d,nd}}$ in an electronic and nuclear contribution and its approximate estimation are in good quantitative agreement with the results from the theoretical simulations and in fair agreement with the experimental results (see Extended Data Fig.~\ref{Fig6ED}).
%Moreover, they allow one to interpret qualitatively the smaller differences in the $A_{2\omega}$ amplitudes observed both in the experiments and in the simulations between the dissociative channel (CH$_3^+$/CD$_3^+$) with respect to the non dissociative one (CH$_4^+$-CD$_4^+$) (see Fig.~\ref{Fig3}c,d and .~\ref{Fig5}c,d).
This analysis clearly indicates that the nuclear dynamics after the XUV photoionisation event plays an important role in determining the channel- and isotope-dependence of the amplitude of oscillations of the sidebands.

In the theoretical model, the phase difference $\Delta\varphi$ between the phases of the sideband oscillations of different isotopic channels originates only from the energy dependence of the electronic part of the matrix dipole moment, as the nuclear contribution given by the functions $N_{\mathrm{H,D}}$ presents only a real part. As a result, the differences are on the order of $\Delta\varphi\approx 0.01$~rad (Fig.~\ref{Fig5}g,h) and are well within the typical error bar ($\approx 0.1$~rad) of our experimental data (Fig.~\ref{Fig3}g,h).

\section*{\label{sec3} Conclusions.}
We have shown that the nuclear dynamics after XUV absorption significantly affects the amplitude of the two-color photoionisation signal in attosecond photoelectron interferometry in molecules. The effects can be interpreted in the spectral domain considering the Fourier transform of the autocorrelation function. The implementation of two isotopologues allows one to highlight the effect of the nuclear dynamics exploiting the different decay times determined by the isotopic substitution. The short duration of the autocorrelation in the time domain (only a few femtoseconds) introduces a finite coherence time for the interaction of the attosecond pulse trains with the molecule. The coherence of the correlated electronic-nuclear wave packets is expected to play a major role in several molecular systems characterized by an ultrafast dynamics after the photoionisation event~\cite{PRL-Vacher-2017} and for the advancement of attosecond metrology~\cite{PRX-Bouchet-2020}.

\subsubsection*{Acknowledgments}

This project has received funding from the European Union's Horizon 2020 research and innovation programme under the Marie Sklodowska-Curie grant agreement no.~641789 MEDEA. D.B. acknowledges support from the Swedish Research Council grant 2020-06384. G.S. acknowledges financial support by the Deutsche Forschungsgemeinschaft project Priority Program 1840 (QUTIF), and grant 429805582 (Project SA 3470/4-1). I.M. and G.S. acknowledge financial support by the BMBF project 05K19VF1, the Deutsche Forschungsgemeinschaft project Research Training Group DynCAM (RTG 2717), and Georg H. Endress Foundation.
D.E. and G.S. acknowledge support and funding by the DFG project International Research Training Group (IRTG) CoCo 2079 and INST 39/1079 (High-Repetition-Rate Attosecond Source for Coincidence Spectroscopy). Z.M. acknowledges support of the PRIMUS (20/SCI/003) project and Czech Science Foundation (20-15548Y). Computational resources were
supplied by the project “e-Infrastruktura CZ” (e-INFRA LM2018140) provided within the program Projects of Large Research, Development and Innovations Infrastructures. This work was also supported by the Ministry of Education, Youth and Sports of the Czech Republic through the e-INFRA CZ (ID:90140).

\subsubsection*{Author Contributions Statement}
D.E, M.S., H.A., M.M. developed the experimental setup. F.F. and L.P. collaborated in the development of the XUV spectrometer. C.D.S., T.P., R.M. and G.S. contributed to the development of the ReMi spectrometer.
D.E, D.B., H.A., M.M. and I.M. performed the experiments. D.E. and D.B analysed the data. J.B. and Z.M developed the numerical simulations for the calculation of the R-Matrix dipole element. S.P. developed the theory and the code for the inclusion of the nuclear dynamics and the influence of the vibrational ground state. G.S. conceived the idea of the experiment, supervised the work and wrote the manuscript, which was discussed and agreed by all coauthors.\\

\subsubsection*{Competing Interests Statement}
The authors declare no competing interests.\\
\subsubsection*{Correspondence and requests for materials}
Correspondence and requests for materials should be addressed to giuseppe.sansone@physik.uni-freiburg.de\\

\clearpage
%\begin{table}
%\centering
%\begin{tabular}{| l | l | l | l | l |}
%\hline
%      &  \multicolumn{2}{c|}{$\mathbf{|c^{(-)}/b|}$} & \multicolumn{2}{c|}{$\mathbf{|c^{(+)}/b|}$} \\ \hline
%      &  \multicolumn{1}{c|}{$\Delta s^{(-)}_R$} &  \multicolumn{1}{c|}{$\Delta \chi^{(-)}$} & \multicolumn{1}{c|}{$\Delta s^{(+)}_R$}  & \multicolumn{1}{c|}{$\Delta \chi^{(+)}$} \\ \hline
%     Exp.    & $0.173\pm0.014$      & $0.155\pm0.008$ & $0.113\pm0.011$   & $0.128\pm0.012$  \\
%     TDSE    & $0.179$              &$0.179$      & $0.122$  & $0.123$    \\
%     SFA     & $0.411$   & $0.411$      & $0.387$  & $0.387$  \\ \hline\hline
%\end{tabular}
%  \caption{Comparison of the ratios ${|c^{(\pm)}/b|}$ obtained from the sinusoidal fits of the variations $\Delta s^{(\pm)}_R$ and $\Delta \chi^{(\pm)}$ according to Eq.~\ref{Eq4} for the experimental data, the TDSE and the SFA simulations.}\label{Table1}
%\end{table}
%\clearpage

\clearpage

%\begin{thebibliography}{10}
%\bibliographystyle{unsrt}
\bibliographystyle{naturemag}
%\printbibliography
%\bibliographystyle{unsrt}
%\bibliography{library.bib}
%\addbibresource{library.bib}
%\end{thebibliography}
%

\clearpage

\section*{Methods}
\setcounter{figure}{0}
\setcounter{table}{0}
%\DeclareCaptionLabelFormat{figure}{\makebox[2.5cm][l]{Fig. #2:}}
%\renewcommand{\thefigure}{Extended Data Fig. \arabic{figure}}
%\renewcommand{\thetable}{\arabic{table} Extended Data}

\subsection*{Methods}
\subsubsection*{Experimental Methods}
Trains of attosecond pulses with photon energies up 50~eV were generated in krypton using a 20~fs driving infrared pulses centered at $\lambda=1012$~nm at 50~kHz repetition rate. The intensity of the field driving the HHG process was estimated in I=10$^{14}$~W/cm$^2$ The temporal delay between the XUV radiation and the IR field was changed in a collinear geometry using a pair of drilled plates~\cite{JP-Ahmadi-2020,Arxiv-Ertel-2022}. The two-color field was focused in the interaction point of the Reaction Microscope using a toroidal mirror operating in one-to-one imaging at an incidence angle of 84$^{\circ}$. The gas target was composed by an equal mixture of CH$_4$ and CD$_4$ molecules. The typical count rate in the measurements was 5-6~kHz and data were acquired for 96~hours. The FWHM of the response function of the REMI photoelectron spectrometer was estimated in $\approx$ 1000~meV. The data discussed in the manuscript were integrated over all emission directions and all orientations of the molecules.

\subsubsection*{Theoretical Methods}
One- and two-photon photoionization matrix elements were calculated using development version of UKRMol+ code~\cite{CPC-Masin-2020} and aug-cc-pVTZ atomic basis set aug-cc-pVTZ for the bound states. The continuum basis consisted of a mixed set of Gaussians reaching to the distance 7.5 atomic units, followed by 10 B-splines spanning the remaining distance to the boundary of the R-matrix inner region at $a = 15$ atomic units. The calculated energy of the ground neutral state was adjusted so that it was exactly 14.4 eV below the calculated energy of the ground ionic state. Bound molecular orbitals were calculated using the 3-state state-averaged complete active space self-consistent field method with restricted open-shell Hartree-Fock reference for CH$_4^+$ in Psi4 v. 1.5~\cite{JCP-Smith-2020}. The lowest three states were equally weighted. Carbon 1s orbitals were frozen in the close-coupling calculations.
The next 11 lowest molecular orbitals were used as active, both for the bound-state complete active space calculation as well as to build the square-integrable part of the close-coupling expansion. The continuum part of the expansion employed 100 cationic states.
Equilibrium geometries and force-fields for the neutral species were calculated at the MP2(fc)/aug-cc-pVTZ level. The quadratic vibronic Hamiltonian for the cation was obtained by diabatizing MR-CIS/CASSCF(7,4) energies of the $^2T_2$ levels in the vicinity of the neutral equilibrium structure. The autocorrelation functions were evaluated for 300 atomic units of time ($\approx 7.26$ fs), correponding to maximum spectral resolution of $\approx 0.3$ eV. Zero-point corrections to the electronic matrix were evaluated using finite-displacements along normal modes, with 7 distorted structures required for each isotopomer. Orientational averaging was performed using order-17 Lebedev grids. Further details of the numerical parameters and procedures are given in Supplementary Information.

%\subsubsection*{Data Availability}
%Experimental data were generated at the FERMI large-scale facility. The experimental and simulations data included in this work are available on the open repository.
%Additional derived data supporting the findings of this study are available from the corresponding author on reasonable request.

%\clearpage
\captionsetup[figure]{name={Extended Data Fig.}}
\captionsetup[table]{name={Extended Data Tab.}}

\begin{figure}[htb]
\centering \resizebox{1.0\hsize}{!}{\includegraphics{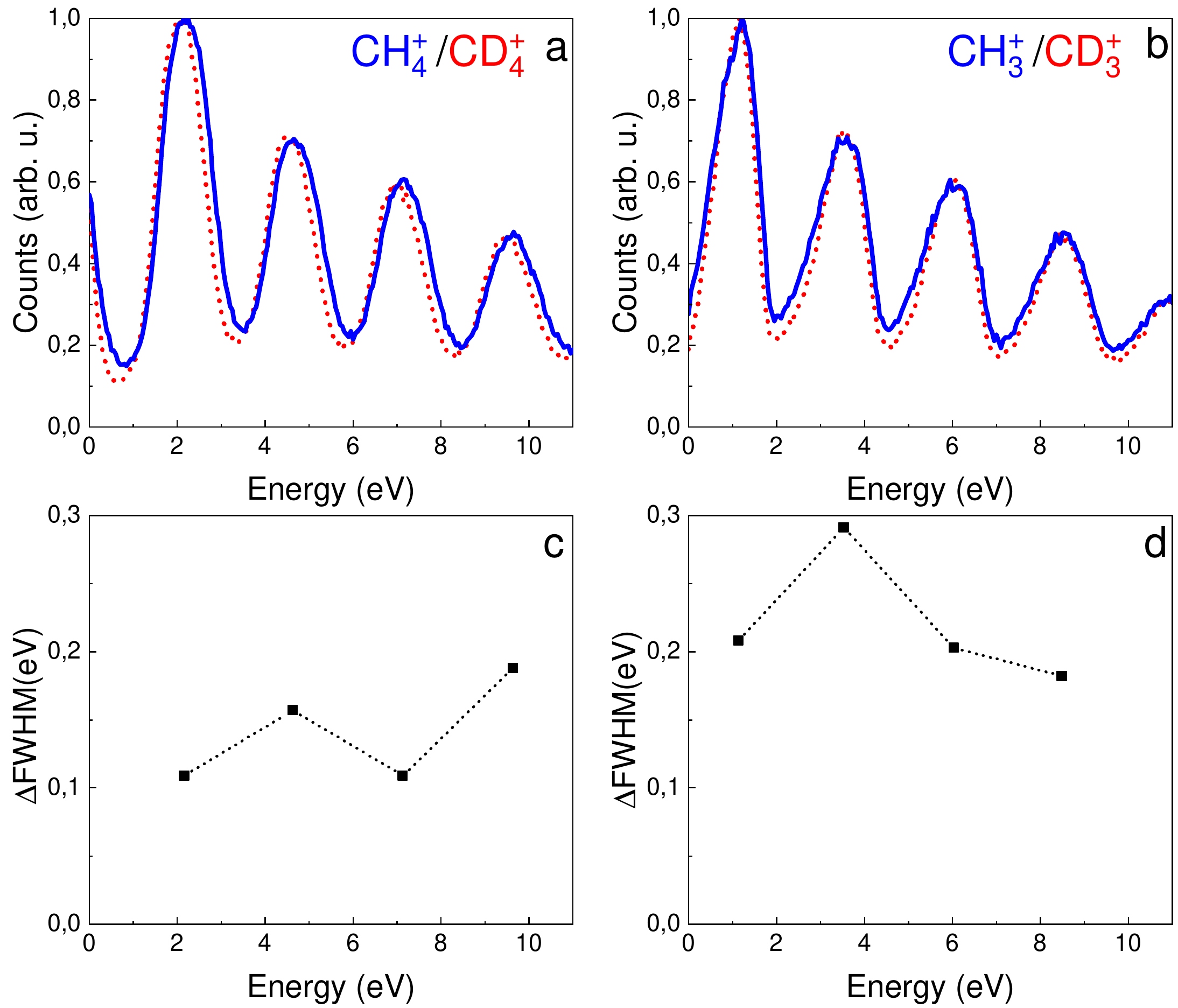}}
\caption{\textbf{Isotopic dependence of channel-resolved XUV-only photoelectron spectra.} Channel-resolved experimental photoelectron spectra measured in coincidence with the ions CH$_4^+$ (blue solid line) and and CD$_4^+$ (red dotted line) (a) and with the ions CH$_3^+$ (blue solid line) and and CD$_3^+$ (red dotted line) (b). Evolution of the differences $\Delta$FWHM=FWHM$_{\mathrm{CH}^+_{3,4}}$-FWHM$_{\mathrm{CD}^+_{3,4}}$ of the FWHM width of the photoelectron peaks measured in coincidence with the ions CH$_4^+$ and CD$_4^+$ (c), and with the ions CH$_3^+$ and CD$_3^+$ (d), respectively.}
\label{Fig1ED}
\end{figure}

%\begin{figure}[htb]
%\centering \resizebox{1.0\hsize}{!}{\includegraphics{Figure/Figure2ED.pdf}}
%\caption{\label{Fig3ED}\textbf{Branching ratios in the photoionisation of methane and deuteromethane.} Branching ratios for the channels CH$_2^+$ (green dotted line), CH$_3^+$ (light blue line), and CH$_4^+$ (blue line), %resulting from the photoionisation of CH$_4$. Photoelectron spectrum resulting from the photoionisation of CH$_4$ (black; data adapted from ref. \cite{}.}
%\end{figure}

\begin{figure}
\centering \resizebox{1.0\hsize}{!}{\includegraphics{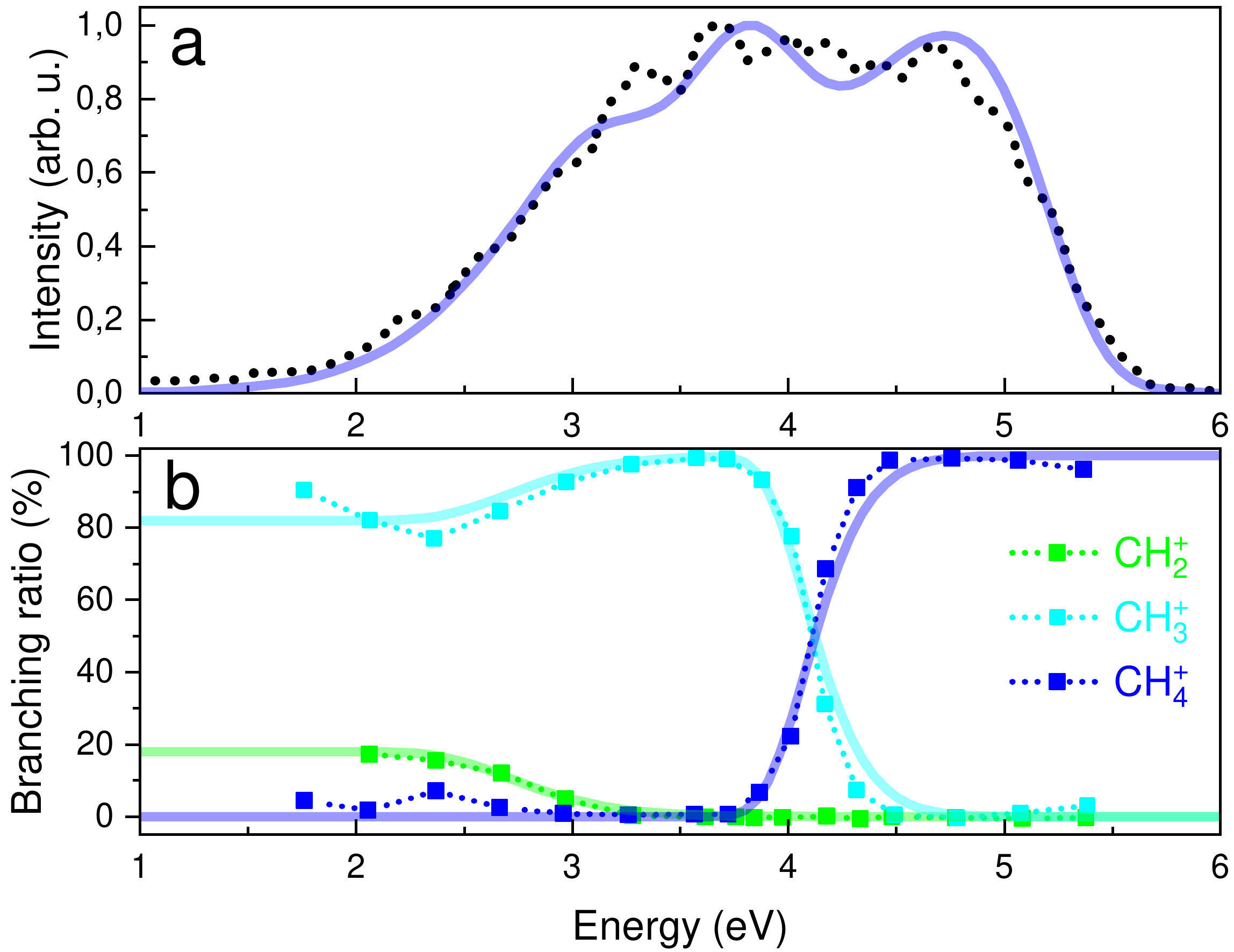}}
\caption{%\textbf{Experimental and simulated photoelectron spectrum and branching ratios in CH$_4$.}
(a) Experimental (black dotted line) and simulated (blue line) photoelectron spectrum generated by the photon energy of the harmonic 15 ($\hbar\omega=18.25$~eV) in CH$_4$. (b) Experimental (square symbols and dotted lines) and adapted (full solid lines) branching ratios for the ionic fragments CH$_2^+$ (green), CH$_3^+$ (light blue), and CH$_4^+$ (blue).
The experimental data taken from ref.~\cite{Field1995} were adapted considering the photon energy difference between the HeI line and the harmonic H15 used in this work.}
\label{Fig2ED}
\end{figure}

\begin{figure}[htb]
\centering \resizebox{1.0\hsize}{!}{\includegraphics{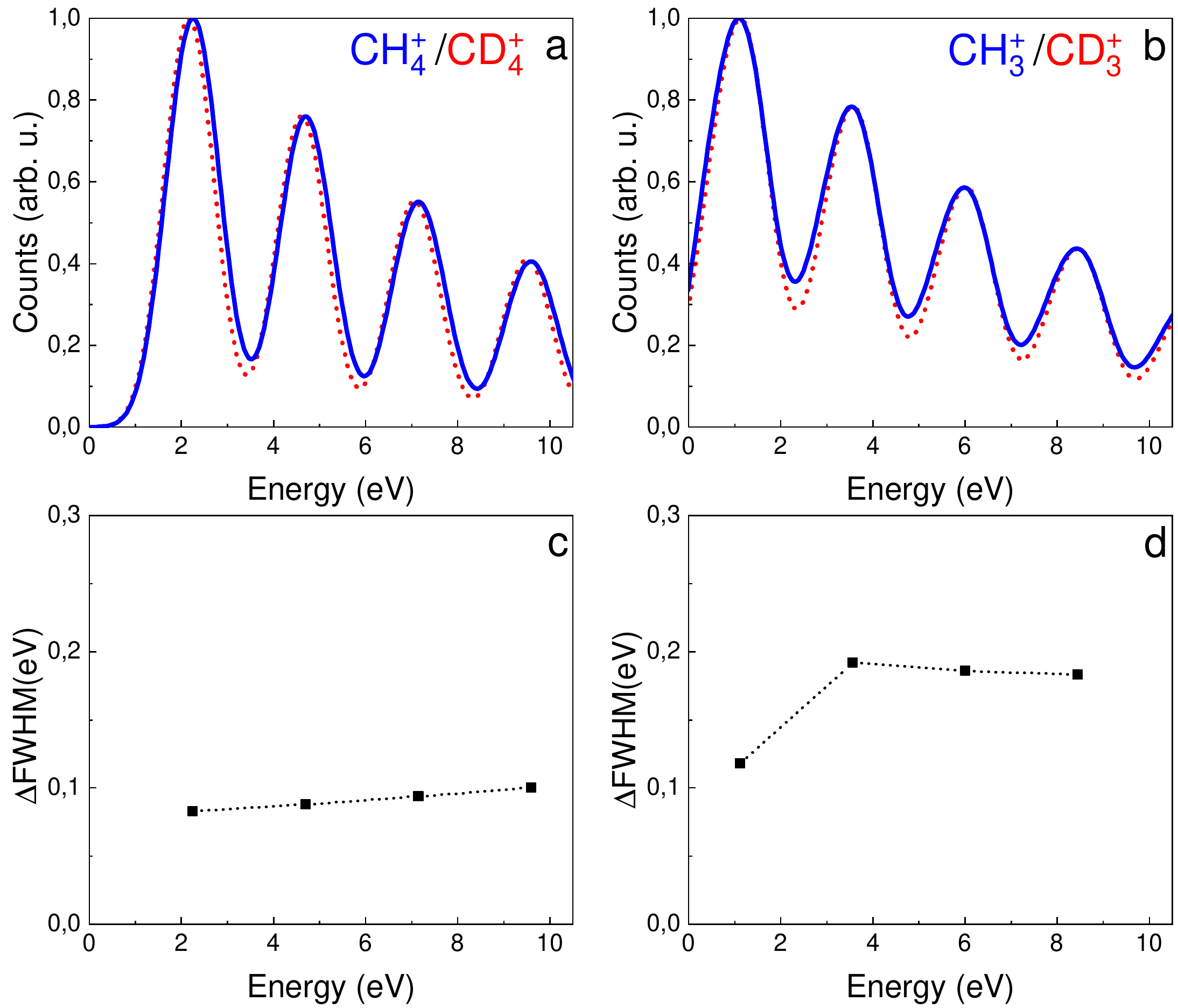}}
\caption{\textbf{Simulations for the isotopic dependence of channel-resolved XUV-only photoelectron spectra.} Channel-resolved simulated photoelectron spectra associated to the ions CH$_4^+$ (blue solid line) and and CD$_4^+$ (red dotted line) (a) and to the ions CH$_3^+$ (blue solid line) and and CD$_3^+$ (red dotted line) (b). Evolution of the differences of the FWHM widths of the simulated photoelectron peaks associated to the ions CH$_4^+$ and CD$_4^+$ (c), and to the ions CH$_3^+$ and CD$_3^+$ (d).}
\label{Fig3ED}
\end{figure}

\begin{figure}[htb]
\centering \resizebox{1.0\hsize}{!}{\includegraphics{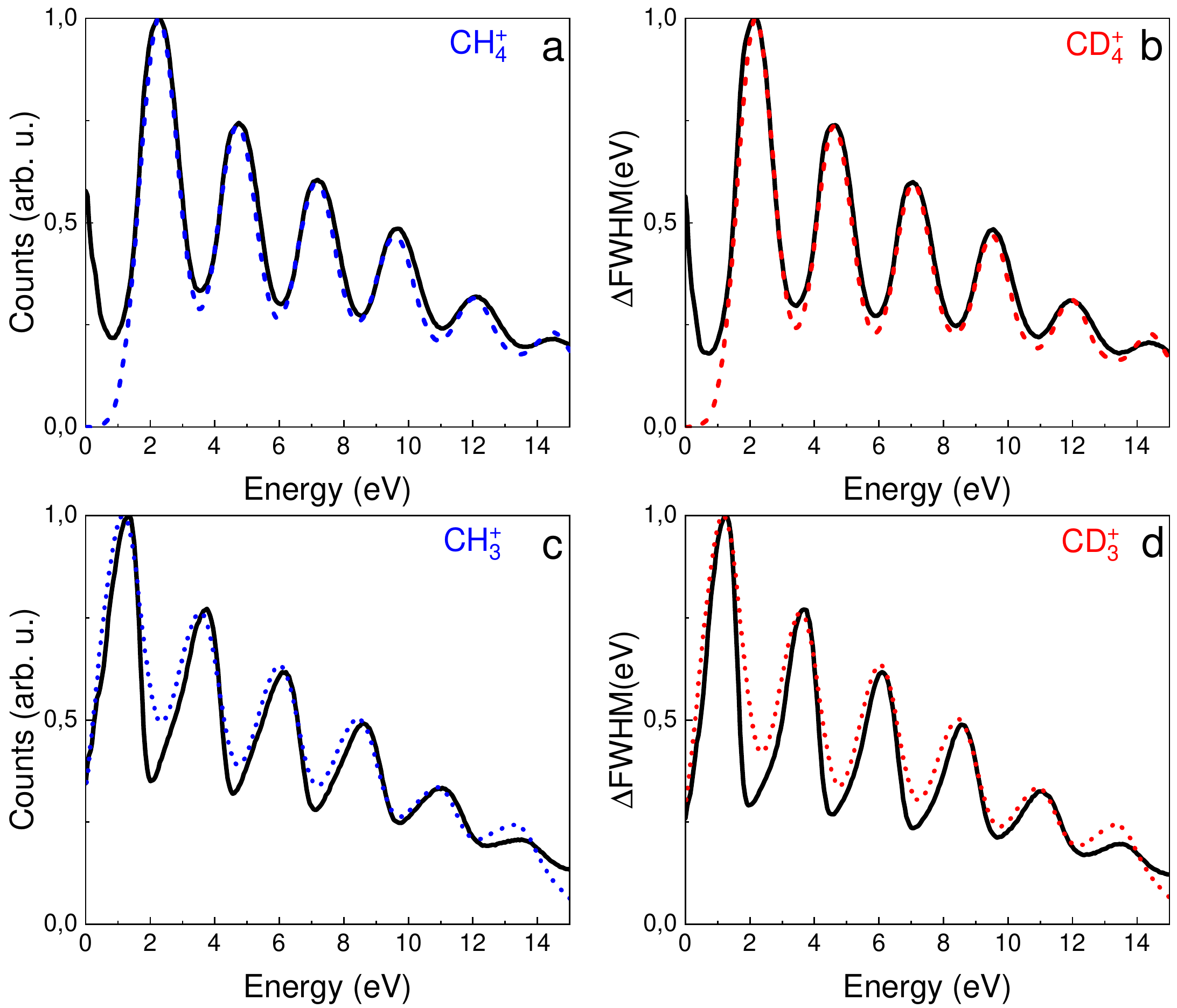}}
\caption{\textbf{Simulated channel-resolved and delay-integrated RABBIT spectrograms in methane and deuteromethane.} Comparison between experimental (black solid lines) and simulated spectra (blue and red lines) measured in coincidence with the ionic fragments CH$_4^+$ (a) (blue dashed line), CH$_3^+$ (b) (blue dotted line), CD$_4^+$ (c) (red dashed line), and CD$_3^+$ (d) (red dotted line).  The experimental and simulated photoelectron spectra were integrated over the delay $\Delta t$ between the attosecond pulse train and the IR pulse.}
\label{Fig4ED}
\end{figure}

\begin{figure}[htb]
\centering \resizebox{1.0\hsize}{!}{\includegraphics{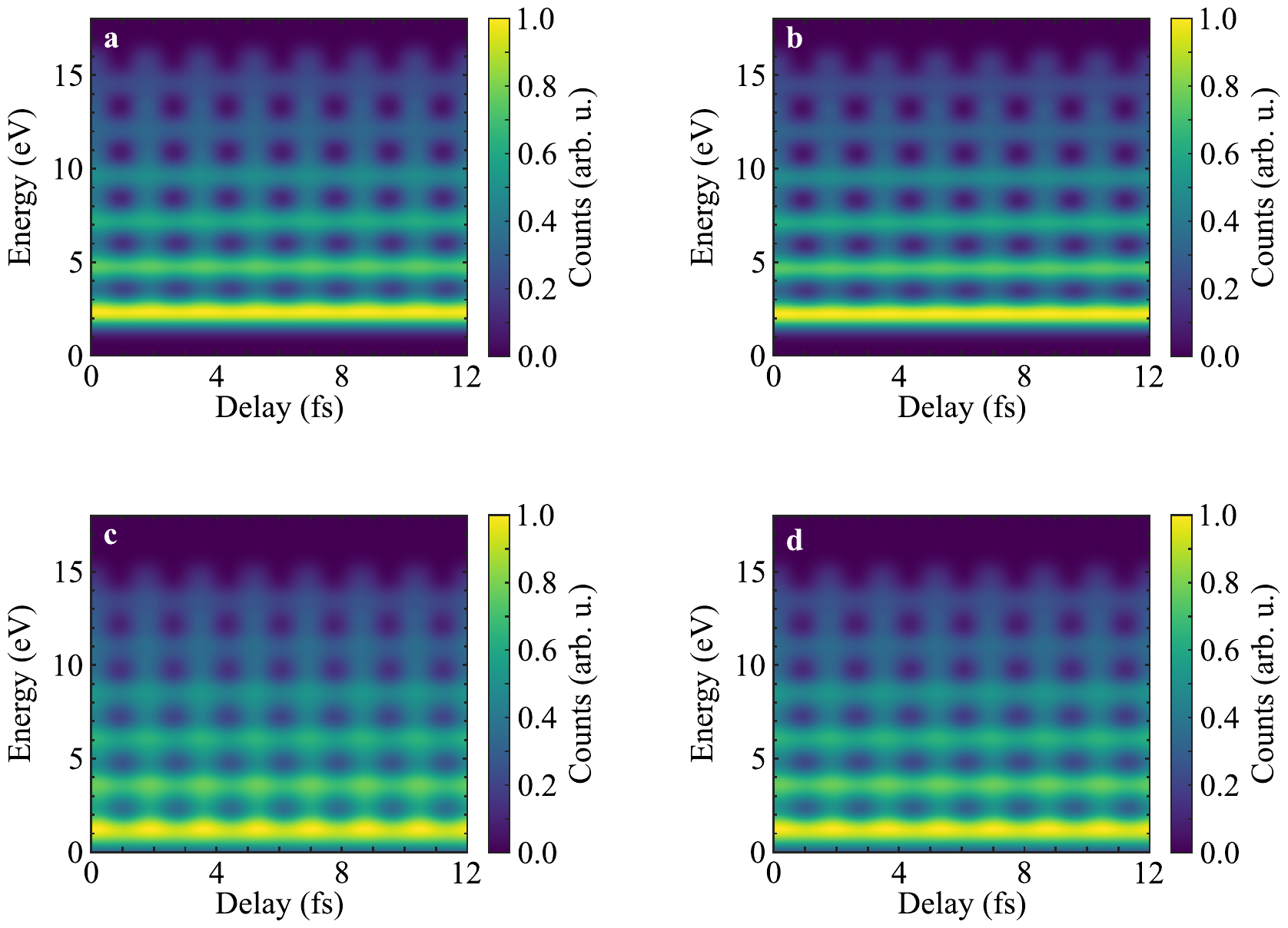}}\caption{\label{Fig6ED}\textbf{Simulated channel-resolved RABBIT spectrograms in methane and deuteromethane.} Simulated RABBIT traces obtained considering only the photoelectrons associated with the ion CH$_4^+$ (a), CD$_4^+$ (b), CH$_3^+$ (c), and CD$_3^+$ (d). In the simulation, the effect of the attosecond chirp was not included.}
\label{Fig5ED}
\end{figure}

\begin{figure}[htb]
\centering \resizebox{1.0\hsize}{!}{\includegraphics{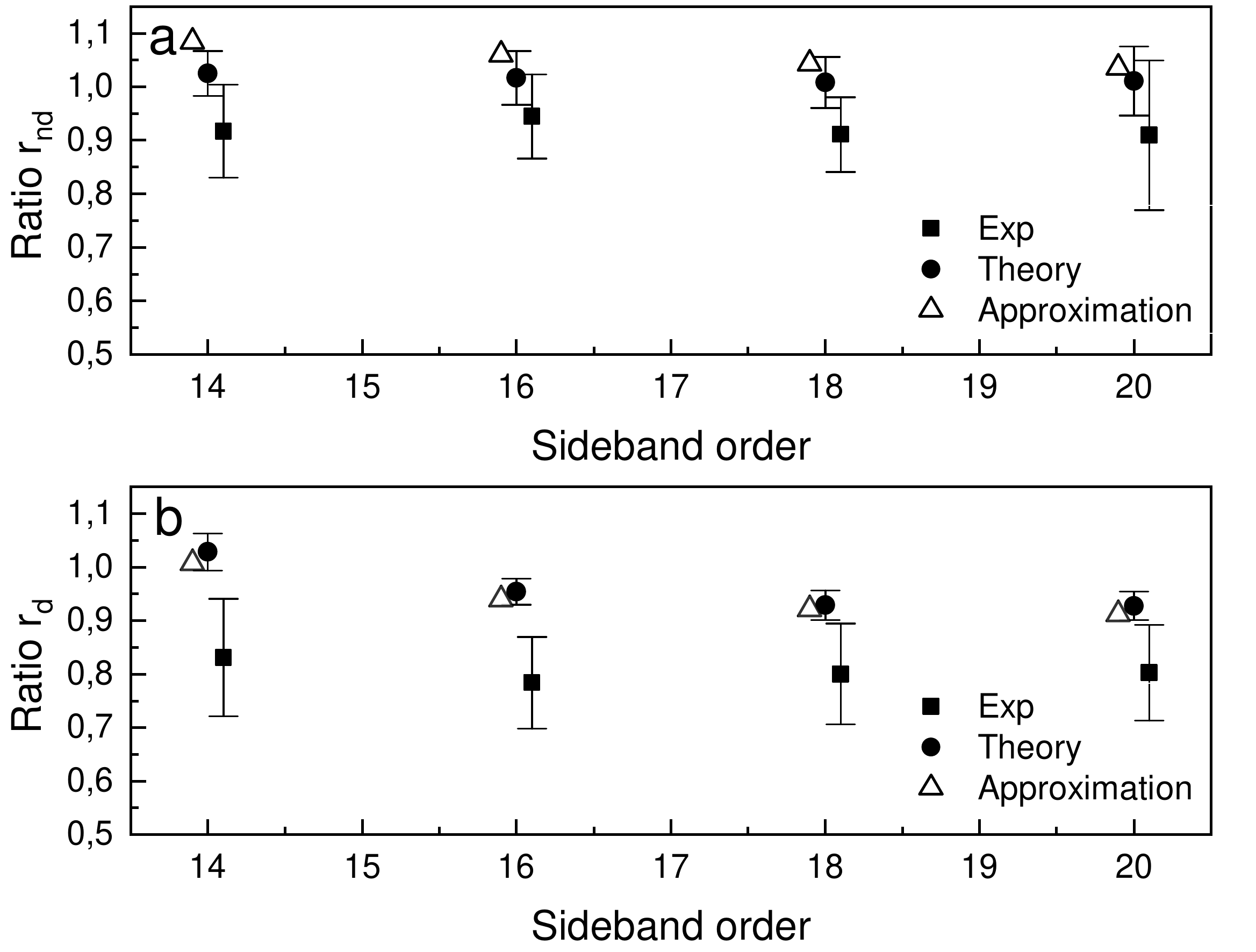}}\caption{\textbf{Ratios of the sideband oscillations for dissociating and non-dissociating channels.} Ratio $\mathrm{r_{nd}}$ (a) and $\mathrm{r_{d}}$ (b) for different sideband orders of the experimental (square) and the theoretical data (circle). The error bars were determined by error propagation of the error bars of the quantities A$_{2\omega}$ shown in Fig.~\ref{Fig3} and Fig.~\ref{Fig5}. The triangles represent the value of the ratios obtained $\mathrm{r_{d,nd}}$ as product of the electronic and nuclear contributions (see Extended Data Table~\ref{Table1}).}
\label{Fig6ED}
\end{figure}

\begin{figure}[htb]
\centering \resizebox{1.0\hsize}{!}{\includegraphics{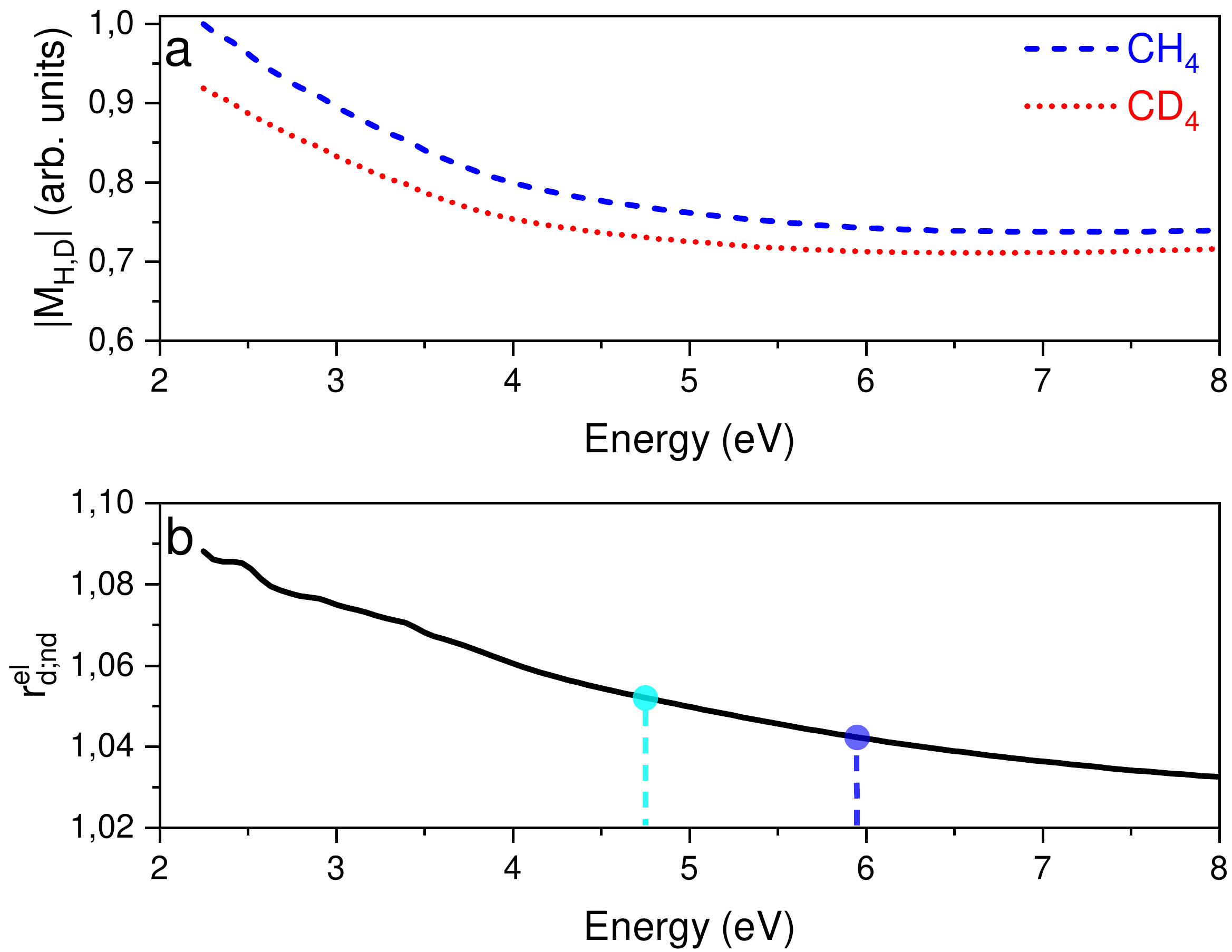}}
\caption{\textbf{Electronic part of matrix dipole moment.} (a) Matrix dipole element for the electronic part for CH$_4$ (blue line) and CD$_4$ (red line) for the energies corresponding to the SB16. (b) Ratio $\mathrm{r^{el}_{d,nd}}$ of the electronic matrix dipole moment for CH$_4$ and CD$_4$ as a function of the photoelectron energy. The light-blue and blue circles indicate the centers of the SB16 associated to the dissociating (CH$^+_3$-CD$^+_3$) and non dissociating channels (CH$^+_4$-CD$^+_4$), respectively.}
\label{Fig7ED}
\end{figure}

\clearpage
\begin{table}
\centering
\begin{tabular}{| l | l | l | l | l |}
\hline
      &  \multicolumn{4}{c|}{Ratio $\mathrm{r_{nd}}$ $(\mathbf{CH_4^+-CD_4^+})$} \\ \hline
     SB order &  \multicolumn{1}{c|}{$14$} &  \multicolumn{1}{c|}{$16$} & \multicolumn{1}{c|}{$18$}  & \multicolumn{1}{c|}{$20$} \\ \hline
     Exp.               & $0.917\pm0.087$      & $0.945\pm0.079$ & $0.911\pm0.069$   & $0.909\pm0.14$  \\
     Theory             & $1.017\pm0.046$     & $1.009\pm0.055$ & $1.000\pm0.052$   & $1.004\pm0.071$  \\
     $\mathrm{r_{nd}^{el}}$      & $1.067$   & $1.042$      & $1.027$  & $1.018$  \\
     $\mathrm{r_{nd}^{nucl}}$    & $1.017$   & $1.017$      & $1.017$  & $1.017$  \\
     $\mathrm{r_{nd}}$           & $1.085$   & $1.060$      & $1.044$  & $1.036$  \\ \hline\hline
           &  \multicolumn{4}{c|}{Ratio $\mathrm{r_d}$ $(\mathbf{CH_3^+-CD_3^+})$} \\ \hline
     SB order &  \multicolumn{1}{c|}{$14$} &  \multicolumn{1}{c|}{$16$} & \multicolumn{1}{c|}{$18$}  & \multicolumn{1}{c|}{$20$} \\ \hline
     Exp.              & $0.831\pm0.109$      & $0.784\pm0.086$ & $0.800\pm0.094$   & $0.802\pm0.089$  \\
     Theory            & $1.023\pm0.033$      & $0.948\pm0.028$ & $0.924\pm0.033$   & $0.923\pm0.029$    \\
     $\mathrm{r_{d}^{el}}$      & $1.128$   & $1.052$      & $1.030$  & $1.020$  \\
     $\mathrm{r_{d}^{nucl}}$    & $0.894$   & $0.894$      & $0.894$  & $0.894$  \\
     $\mathrm{r_d}$             & $1.008$   & $0.940$      & $0.921$  & $0.912$  \\\hline\hline

\end{tabular}
  \caption{Comparison of the ratios $\mathrm{r_{d,nd}}$ obtained from the experimental data (first line) and from the numerical theoretical model (second line). The third and fourth lines present the contribution of the electronic ($\mathrm{r_{d,nd}^{el}}$) and nuclear ($\mathrm{r_{d,nd}^{nucl}}$) ratios to the final value $\mathrm{r_{d,nd}}$=$\mathrm{r_{d,nd}^{el}}\cdot \mathrm{r_{d,nd}^{nucl}}$ (see Eq.~\ref{Eq3}).}\label{Table1}
\end{table}

\clearpage

%\begin{figure}[htb]
%\includegraphics{Figure/Figure4ED.pdf}% Here is how to import EPS art
%\caption{\label{Fig5ED} Channel-resolved photoelectron spectra measured in CH$_4$ (a) (CD$_4$ (b)) in coincidence with the ions CH$_4^+$ (CD$_4^+$) (blue dashed line) and CH$_3^+$ (CD$_3^+$) (red dotted line). The black %solid lines indicates the photoelectron spectra measured in CH$_4$ (a) (CD$_4$ (b)) without selection on the ion species.}
%\end{figure}

\end{document}